\documentclass[prd,aps,nofootinbib,eqsecnum,preprint,superscriptaddress]{revtex4}

\usepackage[latin9]{inputenc}
\usepackage{amsmath,amssymb,bm}
\usepackage[colorlinks]{hyperref}
\usepackage[all]{hypcap}
\usepackage{graphicx,psfrag,float,epsfig}
\usepackage{mathrsfs,wasysym}
\usepackage{epstopdf}
\usepackage{comment}
\usepackage{pbox}
\usepackage{array}
\usepackage{xspace}
\usepackage{feynmf}
\usepackage[usenames,dvipsnames]{color}

\newcommand{\ud}{\mathrm{d}}
\newcommand{\bx}{\mathbf{x}}
\newcommand{\bv}{\mathbf{v}}
\newcommand{\ba}{\mathbf{a}}
\newcommand{\bn}{\mathbf{n}}
\newcommand{\bK}{\mathbf{K}}
\newcommand{\bP}{\mathbf{P}}
\newcommand{\bG}{\mathbf{G}}
\renewcommand{\ud}{d}
\newcommand{\cL}{L}

\newcommand{\refs}{Refs.~}
\newcommand{\refdot}{Ref.~}

\def\addUPitt{Pittsburgh Particle Physics Astrophysics and Cosmology Center, Department of Physics and Astronomy, University of Pittsburgh, Pittsburgh, PA 15260, USA}

\def\addDESY{Deutsches Elektronen-Synchrotron DESY, Notkestrasse 85, 22607 Hamburg, Germany}

\makeatother

\begin{document}

%\title{2PN radiative dynamics of binary systems in the EFT approach}
\title{Radiation Reaction for Non-Spinning Bodies at 4.5PN in the Effective Field Theory Approach}

\author{Adam K.~Leibovich}

\affiliation{\addUPitt}

\author{Brian A.~Pardo}

\affiliation{\addUPitt}

\author{Zixin~Yang}

\affiliation{\addDESY}
\begin{abstract}
We calculate the 2 post-Newtonian correction to the radiation reaction acceleration for non-spinning binary systems, which amounts to the 4.5 post-Newtonian correction to Newtonian acceleration. The calculation is carried out completely using the effective field theory approach. 
The center-of-mass corrections to the results are complicated and are discussed in detail.
Non-trivial consistency checks are performed and we compare with corresponding results in the literature.  Analytic results are supplied in the supplementary materials.

\end{abstract}
\maketitle
%\tableofcontents

%===================================

\section{Introduction}

The tremendous success of the LIGO and VIRGO \cite{LIGOScientific:2014pky,VIRGO:2014yos} gravitational wave detectors and the plans for future, more sensitive, detectors are creating the need for extremely precise theoretical calculations of binary inspirals. During the early stage of the inspiral, it is possible to calculate perturbatively using the post-Newtonian (PN) expansion, which implements an expansion parameter of $v^2/c^2$, where $v$ is the typical relative velocity of the binary constituents. This perturbative expansion is then matched onto numerical results, which are necessary during the late stages of the inspiral phase, due to the breakdown of the PN expansion. With more accurate theoretical calculations of the inspiral, it is potentially possible to extract large amounts of information from the gravitational waveform. 

An effective field theory (EFT) framework named nonrelativistic general relativity (NRGR) \cite{nrgr} has proven to be a useful tool for calculating gravitational wave effects for a binary inspiral. Most of the
calculations in the EFT so far have been in the potential sector, with the state of the art being the 4PN results \cite{nrgr4pn1,nrgr4pn2}, which agree with the
results calculated using other methods \cite{Bini:2013zaa,Damour:2014jta,Bernard:2016wrg}. In the radiation sector, the EFT results 
have recently been calculated to 2PN \cite{radnrgr}, as compared to the 3PN results
calculated using more traditional methods \cite{Blanchet:2001ax}.\footnote{Recent progress has been made in obtaining the dynamics in the non-spinning sector through 2PN \cite{2PN_paper} and through 4PN in the spin sector \cite{pardo,Cho:2021mqw,Cho:2022syn} from the EFT.} However, using the 
EFT result \cite{radnrgr},  in this paper we will calculate the next-to-next-to-leading order (NNLO) non-spinning radiation-reaction
force completely using EFT techniques. This amounts to a 4.5PN correction to the Newtonian acceleration.

Radiation reaction begins at 2.5PN order, first computed  by Burke and Thorne \cite{thorneBT1,thorneBT2}. In the EFT approach, the incorporation of radiation reaction was developed in \refs\cite{chadgsf,chadbr1,Galley:2012qs} by implementing the classical limit of the ``in-in'' approach \cite{inin1,inin2} (see also the formalism developed for nonconservative classical systems in \refs\cite{Galley:2012hx,chadprl2}). At 3.5PN order, the radiation-reaction force was first  calculated in \refs\cite{Iyer1,tail3n,Iyer2,luc96} 
and subsequently rederived using NRGR in \refdot\cite{PhysRevD.86.044029}. The tail effect enters at 4PN order, first calculated in \refs\cite{Damour:2014jta,Bernard:2016wrg} and subsequently in NRGR 
in \refdot\cite{nltail}. The leading spin-orbit and spin-spin effects were calculated in \refs\cite{will1,Will2} in traditional methods and in \refs\cite{natalia1,natalia2} using the EFT. 

In this paper, we calculate the non-spinning radiation-reaction force at 4.5PN order. The paper is organized as follows. In section \ref{sec:EFT}, we review the long-range EFT prescription of NRGR. We also give an overview of nonconservative Lagrangian mechanics that we use to calculate the radiation-reaction diagrams and ultimately derive the equations of motion. Next, in section \ref{sec:NLORR}, we compute the radiation-reaction diagrams term-by-term in the multipole expansion and present the full 4.5PN acceleration in the center-of-mass frame, the main result of this paper. Finally, in section \ref{sec:checks}, we perform consistency checks on our result and compare with corresponding results from the literature, before concluding in \ref{sec:conclude}. We also include an appendix with a discussion of radiative center-of-mass corrections necessary for computing the full result. As the results are somewhat unwieldy, many of the analytical equations are supplied in the supplementary results file. 

Throughout the paper, we will use the total binary mass $m \equiv m_1+m_2$, the mass difference $\delta m=m_1-m_2$, and the reduced mass $\nu \equiv m_1 m_2 / m^2$. We also use following notation for relative coordinates: $\bx^i \equiv \bx_1^i-\bx_2^i \equiv r \bn^i$ as the relative position, $\bv^i \equiv \bv_1^i-\bv_2^i$ the relative velocity, and $\ba^i \equiv \ba_1^i-\ba_2^i$ the relative acceleration.

\section{EFT setup}\label{sec:EFT}

The EFT is used to separate the relevant scales in the binary inspiral by successively integrating out the shorter distance scales, resulting in a hierarchy of EFTs. The successive EFTs are related via matching calculations, which ensure that the long-distance behavior is accurately represented in each system. The short-distance physics is then encapsulated in Wilson coefficients of the operators in the EFT, which are constructed to respect the symmetries of the system (in this case general coordinate invariance).

For the calculation of the radiation reaction, we work with a diffeomorphism invariant effective action, which describes arbitrary gravitational wave sources in the long-wavelength approximation written in terms of multipole moments that live on the binary pair's worldline. In the center-of-mass (COM) frame, the action is given by \cite{andirad}
\begin{equation}\label{Srad}
S_{\rm rad}=-\int dt\sqrt{\bar g_{00}}\biggl[
M(t) - \sum_{\ell=2} \biggl(
\frac1{\ell!} I^L(t) \nabla_{L-2} E_{i_{\ell-1}i_\ell} - \frac{2\ell}{(2\ell+1)!} J^L(t) \nabla_{L-2}B_{i_{\ell-1}i_\ell} 
\biggr)
\biggr],
\end{equation}
where $L = (i_i\cdots i_\ell)$ is a multi-index tensor, and $M(t)$ is the Bondi mass associated with the binary. $I^L(t)$ and $J^L(t)$ are the mass- and current-type source multipole moments, respectively, which depend on the positions $\mathbf x_K$, $K=1,2$, of the massive bodies in the binary. The electric and magnetic components of the Weyl tensor, $E_{ij}$ and $B_{ij}$ depend only on the metric in the radiation region $\bar h_{\mu\nu}$. See \refs\cite{andirad,andirad2} for more details.

\subsection{Calculation of diagrams}
To calculate the nonconservative effects of radiation reaction using the action (\ref{Srad}), we need to formally double the number of degrees of freedom following the approach in \refs\cite{chadgsf,chadbr1}. We take 
\begin{equation}
\mathbf x_K \rightarrow (\mathbf x_{K(1)}, \mathbf x_{K(2)}), \quad \bar h_{\mu\nu} \rightarrow (\bar h_{\mu\nu}^{(1)}, \bar h_{\mu\nu}^{(2)}),
\end{equation}
where the $(1)$ and $(2)$ are the different ``history'' labels of the coordinates and fields. The action is constructed from these degrees of freedom as
\begin{equation}\label{Seff}
    S[\mathbf x_{K(1)},\mathbf x_{K(2)},\bar h_{\mu\nu}^{(1)},\bar h_{\mu\nu}^{(2)}]= S[\mathbf x_{K(1)},\bar h_{\mu\nu}^{(1)}] - S[\mathbf x_{K(2)},\bar h_{\mu\nu}^{(2)}],
\end{equation}
where $S$ includes both the worldline action (\ref{Srad}) and the Einstein--Hilbert action, with appropriate gauge fixing.\footnote{We use linearized harmonic gauge in this work, see \refs\cite{nrgr,radnrgr} for details.} By integrating out the long-wavelength gravitational modes, we obtain the effective action for the open dynamics of the binary inspiral, which can be written as 
\begin{equation}
    S_{\rm eff}[\mathbf x_{K(1,2)}] = \int dt(L[\mathbf x_{K(1)}] - L[\mathbf x_{K(2)}] + R[\mathbf x_{K(1)},\mathbf x_{K(2)}]),
\end{equation}
where $L$ is the usual Lagrangian accounting for the conservative interactions and $R$ is the term containing nonconservative effects. To obtain the radiation-reaction force, we vary the effective action and then take the physical limit, in which the doubled variables are identified with the physical variables, i.e., 
\begin{equation}
    \mathbf x_{K(1)},\mathbf x_{K(2)} \to \mathbf x_{K}.
\end{equation}
In this work, we make a convenient coordinate redefinition to ``plus-minus'' coordinates, defined by
\begin{equation}
    \mathbf{x}_{K+} \equiv(\mathbf{x}_{K(1)}+\mathbf{x}_{K(2)}) / 2, \quad \mathbf{x}_{K-} \equiv \mathbf{x}_{K(1)}-\mathbf{x}_{K(2)},
\end{equation}
which simplifies the procedure for deriving the dynamics.
We can then vary the action with respect to the minus degrees of freedom, and then take the physical limit by simply setting
\begin{equation}
    \mathbf{x}_{K+}\rightarrow\mathbf{x}_K,\quad \mathbf{x}_{K-}\rightarrow0.
\end{equation}
The nonconservative acceleration is then given by varying $R$ in Eq.~\eqref{Seff} as
\begin{equation}
\ba_K^i(t)=\frac1{m_K}\frac{\delta R}{\delta \bx^i_{K-}\!}-
\frac1{m_K}\frac{d}{d t}\frac{\delta R}{\delta \bv^i_{K-}\!}
+\cdots
\bigg|_{\substack{\mathbf {x}_{K-}\rightarrow0 \\ \mathbf {x}_{K+}\rightarrow\mathbf {x}_K}}.\label{eq:variation}
\end{equation}

The topologies that we need to consider up to 4.5PN order are given by%
\footnote{Note that we only have to consider diagrams to linear order in the radiation modes, neglecting nonlinear radiative effects that only contribute starting at the 5PN order. It is worth observing, however, that nonlinear gravitational effects enter through the multipole moments themselves within the potential regime. Furthermore, there is a new topology at 4PN that accounts for the tail term, which we do not consider here. See \refdot\cite{nltail} for details.}%
\begin{equation}
iS_\mathrm{eff}[\mathbf x_K^\pm]\quad=iS_\mathrm{con}\quad+\quad\sum_{l\geq2}\quad
  \parbox{50pt}{
    \includegraphics{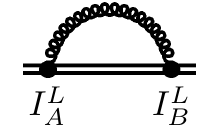}
  }
  \qquad+\quad
  \parbox{50pt}{\includegraphics{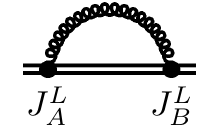}
  }\quad.
\end{equation}
We consider the contributions from the conservative action, $S_{\mathrm {con}}$, separately, focusing now on the dissipative terms. For this topology, we can find a general expression in terms of multipole moments and their derivatives
given by \cite{nltail}
\begin{equation}
\int dt\,R =\sum_{\ell\geq2}\frac{(-1)^{\ell+1}(\ell+2)G}{(\ell-1)}\int dt\,\biggl[\frac{2^{\ell}(\ell+1)}{\ell(2\ell+1)!}I_{-}^{L}(t)I_{+}^{L(2\ell+1)}(t)+\frac{2^{\ell+3}\ell}{(2\ell+2)!}J_{-}^{L}(t)J_{+}^{L(2\ell+1)}(t)\biggr],\label{eq:action}
\end{equation}
where $I_{-}^L \equiv I_{(1)}^L-I_{(2)}^L$ and $I_{+}^L \equiv(I_{(1)}^L+I_{(2)}^L) / 2$, and  $I_A^L \text { for } A=(1),(2)$ are the different history versions of the mass-type multipoles, with similar expressions for the current-type multipoles.

The leading order nonconservative acceleration, entering at 2.5PN, can be computed from the mass-quadrupole component of Eq.~\eqref{eq:action} given by $R_{2.5\mathrm{PN}}=-\frac{G}{5}I_{-}^{ij}I_{+}^{ij(5)}$ using Eq.~\eqref{eq:variation}; the result is simply the usual Burke--Thorne equation
\begin{equation}
    \ba_K^i= -\frac{2G}{5}\bx^j_K I^{ij(5)}_\mathrm{0PN}.\label{eq:aBT}
\end{equation}
Similarly, the 3.5PN acceleration was computed in \refdot\cite{PhysRevD.86.044029}. 

\section{Radiation reaction through 4.5PN}\label{sec:NLORR}
During the inspiral phase, the relative velocity $v$ of the bodies is small, so we can expand the acceleration in the PN expansion. For non-spinning binaries, we have
\begin{equation}
\ba^i = \ba_{0\textrm{PN}}^i + \ba_{1\textrm{PN}}^i + \ba_{2\textrm{PN}}^i + \ba_{2.5\textrm{PN}}^i + \ba_{3\textrm{PN}}^i + \ba_{3.5\textrm{PN}}^i + \ba_{4\textrm{PN}}^i + \ba_{4.5\textrm{PN}}^i + \cdots,
\end{equation}
where the subscript denotes the PN order of the term. The leading term,  $\ba_{0\textrm{PN}}^i$, is just the Newtonian acceleration. The 1PN correction is the Einstein--Infeld--Hoffmann correction, which scales as $\mathcal O(v^2)$, while the 2PN correction scales as $\mathcal O(v^4)$. At 2.5PN order, we have the leading order radiation reaction, or Burke--Thorne, term. This is the first nonconservative piece of the acceleration. At 3PN order, we again have a conservative correction. At 3.5PN, we have the first correction to the Burke--Thorne term. At 4PN, we have a mix between conservative and nonconservative contributions, including the leading tail contribution. These contributions have all been calculated using traditional methods \cite{Bini:2013zaa,Damour:2014jta,Bernard:2016wrg,thorneBT1,thorneBT2,Iyer1,tail3n,Iyer2,luc96} and using the EFT approach \cite{nrgr4pn1,nrgr4pn2,chadgsf,chadbr1,PhysRevD.86.044029,nltail}. Finally, at 4.5PN we get the 2PN correction to the Burke--Thorne term, which is the focus of this paper.

In this section, we compute the radiation-reaction equations of motion at 4.5PN order. For simplicity, we break the calculation into distinct multipole terms as well as contributions from the conservative sector arising from order-reduced accelerations. 
Explicitly, through the 4.5PN order, the action that yields the radiation-reaction
acceleration can be written as
\begin{equation}
R_{4.5\mathrm{PN}}=-\frac{G}{5}I_{-}^{ij}I_{+}^{ij(5)}-\frac{16G}{45}J_{-}^{ij}J_{+}^{ij(5)}+\frac{G}{189}I_{-}^{ijk}I_{+}^{ijk(7)}+\frac{G}{84}J_{-}^{ijk}J_{+}^{ijk(7)}-\frac{G}{9072}I_{-}^{ijkl}I_{+}^{ijkl(9)},
\label{eq:lagrangian}
\end{equation}
where the first term enters at 2.5PN and the first three terms contribute up to order 3.5PN. Due to PN corrections to these multipoles, these terms also contribute at 4.5PN. We now proceed to compute the 4.5PN acceleration term by term.

\subsection{Mass quadrupole}\label{sec:Amq}

The mass quadrupole can be expanded as%
\footnote{We neglect spin dependence in the multipole moments, which enter here at 1.5PN for spin-orbit couplings \cite{amps} and 2PN for spin-spin couplings \cite{Cho:2021mqw, Cho:2022syn, Porto:2010zg}. }%
\begin{align}
I^{ij}=&I_\mathrm{0PN}^{ij}+\epsilon I_\mathrm{1PN}^{ij}+\epsilon^{2}I_\mathrm{2PN}^{ij}+\mathcal O(\epsilon^{2.5})\nonumber\\
=& \sum_{K\neq L} m_K\biggl\{\bx_K^i \bx_K^j+\epsilon\biggl[\biggl(\frac{3}{2} \bv_K^2-\sum_{L \neq K} \frac{G m_L}{r}\biggr) \bx_K^i \bx_K^j\nonumber\\
&+\frac{11}{42} \frac{d^2}{d t^2}(\bx_K^2 \bx_K^i \bx_K^j)-\frac{4}{3} \frac{d}{d t}(\bx_K \cdot \bv_K \bx_K^i \bx_K^j)\biggr]\biggr\}_{\mathrm{TF}} + \mathcal O(\epsilon^2),
\end{align}
where $\epsilon$ counts the PN order and the $\mathcal O(\epsilon^2)$ expression can be found in \refdot\cite{radnrgr}.
Then through 4.5PN, the mass quadrupole component of the action can be written as
\begin{align}
S_{\text{mq}}=&-\frac{G}{5}\int dt\,[I_{0-}^{ij}I_{0+}^{ij(5)}+\epsilon(I_{0-}^{ij}I_{1+}^{ij(5)}+I_{1-}^{ij}I_{0+}^{ij(5)})
\nonumber\\
&\qquad+\epsilon^{2}(I_{0-}^{ij}I_{2+}^{ij(5)}+I_{2-}^{ij}I_{0+}^{ij(5)}+I_{1-}^{ij}I_{1+}^{ij(5)})+\cdots],
\end{align}
where the numeric subscript on the multipole moments is its PN order.
The $\mathcal O(\epsilon^{0})$ and $\mathcal O(\epsilon^{1})$ terms correspond to
the 2.5PN and 3.5PN radiation-reaction mass quadrupole contributions, respectively. However, upon variation, these terms also contribute to the 4.5PN acceleration through order reduction of accelerations.
Looking specifically at the terms that depend on $I^{ij}_{0-}$, after varying the action, we find
\begin{equation}
    \ba_K^i = -\frac{2G}{5}\bx^j_K (I^{ij(5)}_\mathrm{0}+\epsilon I_1^{ij(5)}+\epsilon^2I_2^{ij(5)}).
\end{equation}
Each of these terms contributes at 4.5PN, as each of these terms is itself dependent on accelerations and higher time derivatives that are not of definite PN order. The first term receives corrections through order reduction using the 2PN conservative acceleration or two order-reduced 1PN accelerations. The second term has corrections from the 1PN acceleration. The final term only requires order reduction using the Newtonian acceleration. We must of course also vary the terms that depend upon $I_{1-}^{ij}$ and $I_{2-}^{ij}$. These are more complicated, but follow similarly to the above. For instance, for the $I_{1-}^{ij}I_{0+}^{ij(5)}$ term, one will again need the order-reduced 1PN acceleration.

In this general frame, this completes the 4.5PN mass quadrupole contribution. However, we must consider additional terms that arise when making the coordinate transformation to the COM frame, in which we compute the relative acceleration $\ba^i = \ba_{1}^i-\ba_{2}^i$. The coordinate shift can be written as
\begin{align}
\bx^i_{1}=&\frac{m_2}{m} \bx^i+\delta \bx^i, \\
\bx^i_{2}=&-\frac{m_1}{m} \bx^i+\delta \bx^i,
\end{align}
where $\delta \bx^i$ is given order by order in a PN expansion as
\begin{equation}
    \delta \bx^i = \epsilon \delta \bx^i_{1\textrm{PN}}+\epsilon^2 \delta \bx^i_{2\textrm{PN}}+\epsilon^{2.5} \delta \bx^i_{2.5\textrm{PN}}+\epsilon^3\delta \bx^i_{3\textrm{PN}}+\epsilon^{3.5} \delta \bx^i_{3.5\textrm{PN}}+\mathcal O(\epsilon^4).
\end{equation}
The terms $\delta \bx^i_{2\textrm{PN}}$ was recently computed using EFT methods in \refdot\cite{radnrgr}, while $\delta \bx^i_{2.5\textrm{PN}}$ vanishes in our gauge.\footnote{Contrast this with the nonzero COM correction at 2.5PN computed in \refdot\cite{Blanchet:2002mb} using harmonic coordinates.} The 3PN COM shift has yet to be computed in our gauge, but is beyond the scope of this paper as it does not contribute at the 4.5PN order in the dissipative sector. The relevant terms, which we reproduce here for convenience, are
\begin{align}
\delta \bx_{1\textrm{PN}}^i &=\frac{\nu \delta m}{2 m} \bx^i\biggl(\bv^2-\frac{G m}{r}\biggr), \\
\delta \bx_{2\textrm{PN}}^i &=\frac{\nu \delta m}{2 m}\biggl\{\bx^i \biggl[\biggl(\frac{3}{4}-3 \nu\biggr) 
\bv^4+\frac{G m}{r}\biggl(\biggl(\frac{19}{4}+3 \nu\biggr) \bv^2\biggr)\nonumber\\
&\qquad+\biggl(-\frac{1}{4}+\frac{3 \nu}{2}\biggr) \dot{r}^2+\biggl(\frac{7}{2}-\nu\biggr) 
\frac{G m}{r}\biggr]-\bv^i\biggl[\frac{7}{2} G m \dot{r}\biggr]\biggr\}.
\end{align}
Note that $\delta \bx^i_{3.5\textrm{PN}}$ is nonvanishing in our gauge (see Appendix \ref{sec:appendix}), and will be discussed in section \ref{sec:Aco} as it contributes at 4.5PN when shifting to the COM frame within the conservative sector.

We apply these COM coordinate transformations to the 2.5PN and 3.5PN mass quadrupole terms in the acceleration in relative coordinates. Working with the multipole moments in the COM frame, we have a contribution given by
\begin{equation}
    \ba^i_\mathrm{COM}=-\frac{2G}5\bx^j\frac{d^5}{dt^5}(m\delta \bx_{1\textrm{PN}}^i\delta \bx_{1\textrm{PN}}^j-\frac13m\delta \bx_{1\textrm{PN}}^2 \delta^{ij}),
\end{equation}
where the 2PN shift does not contribute due to the symmetry of the mass quadrupole moment. Additionally, we find that there will be COM corrections to expressions containing the 1PN multipole moment when applied after variation with respect to the minus coordinates. Adding these corrections to our result yields a final expression for the COM frame mass quadrupole contribution, which can be found in the supplemental file.

\subsection{Mass octupole}

The mass octupole can be written as
\begin{align}
I^{ijk}&=I_{0}^{ijk}+\epsilon I_{1}^{ijk}+\mathcal O(\epsilon^2)\\
&=\sum_{A\neq B} m_A\biggl\{\biggl[\biggl(1+\frac{3}{2} \bv_A^2-\sum_{B \neq A} \frac{Gm_B}{r}\biggr) \bx_A^i \bx_A^j \bx_A^k+\frac{1}{18} \frac{d^2}{d t^2}(\bx_A^2 \bx_A^i \bx_A^j \bx_A^k)\biggr]_{\mathrm{STF}}\nonumber\\
&\qquad-\frac{7}{9} \frac{d}{d t}\bigl[(\bx_A^i \bx_A^j \bx_A^k \bx_A^l)_{\mathrm{STF}} \bv^l\bigr]\biggr\}+\mathcal O(\epsilon^2)
,
\end{align}
again neglecting spin.
Then the octupole contributions to the action, using Eq.~(\ref{eq:lagrangian}), are
\begin{equation}
S_{\text{mo}}=\frac{G}{189}\int dt\,[\epsilon I_{0-}^{ijk}I_{0+}^{ijk(7)}+\epsilon^{2}(I_{1-}^{ijk}I_{0+}^{ij(7)}+I_{0-}^{ijk}I_{1+}^{ij(7)})+\cdots].\label{eq:oct-action}
\end{equation}
The $\mathcal O(\epsilon)$ term contributes to the 3.5PN radiation-reaction
acceleration, as well as to the 4.5PN acceleration upon reducing accelerations using the 1PN acceleration. 
%The $\mathcal O(\epsilon^{2})$ terms, then, are used to compute the4.5PN acceleration. 
As an example, after varying the third term, proportional to $I_{0-}^{ijk}I_{1+}^{ij(7)}$, we find that
\begin{equation}
    \ba^i_1 = \frac{G}{63}\bx^j_1\bx^k_1I_1^{ijk(7)}.
\end{equation}
We additionally must consider the first term in Eq.~(\ref{eq:oct-action}) with 1PN acceleration reductions and the second term with Newtonian acceleration, which together with the above yield the 4.5PN acceleration. This completes the octupole term in the original frame. When transforming to the COM frame, we pick up an additional 4.5PN piece from the 1PN COM shift in the 3.5PN term. The mass octupole expression for $\ba^i$ in the COM frame can be found in the supplementary file.

\subsection{Current quadrupole}

The current quadrupole can be written as
\begin{align}
J^{ij}=&\epsilon^{0.5}J_{0}^{ij}+\epsilon^{1.5}J_{1}^{ij}+\mathcal O(\epsilon^{2.5})\\
=&
\sum_A m_A\biggl(1+\frac{\bv_A^2}{2}\biggr)\bigl[(\bx_A \times \bv_A)^i \bx_A^j\bigr]_{\mathrm{STF}}
\nonumber\\&+\sum_{A\neq B} \frac{G m_A m_B}{r}\biggl[2(\bx_A \times \bv_A)^i \bx_A^j 
-\frac{11}{4}(\bx_B \times \bv_A)^i \bx_B^j-\frac{3}{4}(\bx_B \times \bv_A)^i \bx_A^j
\nonumber \\
&\qquad+(\bx_A \times \bv_A)^i \bx_B^j+\frac{7}{4}(\bx_A \times \bx_B)^i \bv_A^j
+\frac{\bv_A \cdot \bx}{4 r^2}(\bx_A \times \bx_B)^i(\bx_A^j+\bx_B^j)\biggr]_{\mathrm{STF}}
\nonumber\\
&+\frac{1}{28} \frac{d}{d t}\biggl[\sum_A m_A(\bx_A \times \bv_A)^i(3 \bx_A^2 \bv_A^j-\bx_A \cdot \bv_A \bx_A^j) 
\nonumber\\
&\qquad+\sum_{A\neq B} \frac{G m_A m_B}{2 r^3} \bx_A^i(\bx_A \times \bx_B)^j(6 \bx_A^2-7 \bx_A \cdot \bx_B+7 \bx_B^2)\biggr]_{\mathrm{STF}} +\mathcal O(\epsilon^{2.5}),
\end{align}
and thus the term in the action that contributes through 4.5PN is given by
\begin{equation}
S_{\text{cq}}=-\frac{16G}{45}\int dt\,[\epsilon J_{0-}^{ij}J_{0+}^{ij(5)}+\epsilon^{2}(J_{1-}^{ij}J_{0+}^{ij(5)}+J_{0-}^{ij}J_{1+}^{ij(5)})+\cdots].\label{eq:cquad-action}
\end{equation}
The current quadrupole term first enters the acceleration at 3.5PN through the first term in Eq.~(\ref{eq:cquad-action}). The 4.5PN acceleration contains three contributions: two pieces from the 1PN current quadrupole in either the plus or minus coordinates, and a third from reducing accelerations in the 3.5PN acceleration with the 1PN conservative acceleration. Upon shifting to the frame of the COM, we find an additional contribution from the 1PN coordinate correction in the 3.5PN acceleration. The full expression can be found in the supplementary file.

\subsection{Mass hexadecapole}

The mass hexadecapole term contributes first at the 4.5PN order and
is given by
\begin{equation}
S_{\text{mh}}=-\frac{G}{9072}\int dt\,\epsilon^{2}I_{0-}^{ijkl}I_{0+}^{ijkl(9)}+\mathcal O(\epsilon^3).
\end{equation}
Thus, we only need the leading order expression of the mass hexadecapole,
given by
\begin{equation}
I_{0}^{ijkl}=\sum_{A}m_{A}[\bx_{A}^{i}\bx_{A}^{j}\bx_{A}^{k}\bx_{A}^{l}]_{\text{STF}}.
\end{equation}
Upon variation, we find that
\begin{equation}
\ba_{1\text{mh}}^{i}=-\frac{G}{2268}\bx_1^{j}\bx_1^{k}\bx_1^{l}I_{0}^{ijkl(9)},
\end{equation}
and in the COM frame 
\begin{equation}
\ba_{\text{mh}}^{i}=-\frac{G}{2268}(1-3\nu)\bx^{j}\bx^{k}\bx^{l}I_{0}^{ijkl(9)},
\end{equation}
which can be found in the supplementary file.

\subsection{Current octupole}\label{sec:Aco}

The current octupole term contributes first at the 4.5PN order as
well and is given by
\begin{equation}
S_{\text{co}}=\frac{G}{84}\int dt\,\epsilon^{2}J_{0-}^{ijk}J_{0+}^{ijk(7)} +\mathcal O(\epsilon^3).
\end{equation}
Thus, we only need the leading order expression of the current octupole,
given by
\begin{equation}
J_{0}^{ijk}=\sum_{A}m_{A}[\epsilon^{ilm}\bx_{A}^{j}\bx_{A}^{k}\bx_{A}^{l}\bv_{A}^{m}]_{\text{STF}}=m\nu(1-3\nu)[\epsilon^{ilm}\bx^{j}\bx^{k}\bx^{l}\bv^{m}]_{\text{STF}}.
\end{equation}
Upon variation, we find that the contribution to the acceleration
from the current octupole is
\begin{equation}
\ba_{1\text{co}}^{i}=\frac{G}{84}\bx_1^{j}\bx_1^{k}[2(\epsilon^{ilm}J_{0}^{jkm(7)}+\epsilon^{ijm}J_{0}^{klm(7)}+\epsilon^{jlm}J_{0}^{ikm(7)})\bv_1^{l}+\epsilon^{ijm}\bx_1^{l}J_{0}^{klm(8)}],
\end{equation}
and in relative coordinates
\begin{equation}
\ba_{\text{co}}^{i}=\frac{G}{84}(1-3\nu)\bx^{j}\bx^{k}[2(\epsilon^{ilm}J_{0}^{jkm(7)}+\epsilon^{ijm}J_{0}^{klm(7)}+\epsilon^{jlm}J_{0}^{ikm(7)})\bv^{l}+\epsilon^{ijm}\bx^{l}J_{0}^{klm(8)}].
\end{equation}
Again, the COM contribution can be found in the supplementary file.

\subsection{Conservative acceleration reductions}

In this section, we discuss 4.5PN terms that result from corrections to the conservative accelerations at lower orders. Variation of the conservative Lagrangian yields accelerations that are themselves acceleration dependent. Order reducing these conservative terms with nonconservative accelerations yields additional nonconservative corrections at 4.5PN. Additionally, since the COM is no longer conserved at 3.5PN, there arise nonconservative corrections at 4.5PN in relative coordinates when shifting to the COM frame.

In the 1PN acceleration, we obtain a reduced contribution from inserting
the 3.5PN acceleration as
\begin{equation}
\ba_{1}^{\mathrm{(red)}}=\biggl[\frac{1}{2}\frac{Gm_{2}}{r}(\ba_{2}\cdot \bn)\bn^{i}-(\ba_{1}\cdot \bv_{1})\bv_{1}^{i}-\ba_{1}^{i}\biggl(3\frac{Gm_{2}}{r}+\frac{1}{2}\bv_{1}^{2}\biggr)+\frac{7}{2}\frac{Gm_{2}}{r}\ba_{2}^{i}\biggr]_{\ba_{\text{3.5PN}}}.
\end{equation}
The full 3.5PN acceleration can be found in the supplementary materials. There is an additional piece resulting from order reducing accelerations and higher coordinate derivatives in the 2PN acceleration using the leading order Burke--Thorne acceleration at 2.5PN,  yielding a 4.5PN correction. This concludes the calculation of the 4.5PN acceleration in the original coordinate system.

When shifting to relative coordinates in the COM frame, we must consider one additional contribution. Naively, we would expect to have a 2.5PN COM correction to the 2PN acceleration; however, this vanishes because there is no 2.5PN COM shift in our gauge. However, there is a nonzero 3.5PN correction, as discussed in Appendix \ref{sec:appendix}. Applying this to the 1PN acceleration yields
\begin{align}
\ba_{\text{COM}}^{\text{i}} =\frac{\delta m}{m}\biggl[&\biggl(-\frac{G}{2r}+\frac{1}{2}\bv^{2}\biggr)\delta\ddot{\bx}_{3.5\textrm{PN}}^{i}
-\frac{Gm}{2r^{3}}(\bx\cdot\delta\ddot{\bx}_{3.5\textrm{PN}})\bx^{i}+(\bv\cdot\delta\ddot{r}_{3.5\textrm{PN}})\bv^{i}\nonumber\\
&-\frac{Gm}{r^{3}}\biggl(2(\bx\cdot\delta\dot{\bx}_{3.5\textrm{PN}})\bv^{i}+(\bv\cdot\delta\dot{\bx}_{3.5\textrm{PN}})\bx^{i}\biggr)\biggr],
\end{align}
where $\delta \bx_{3.5\textrm{PN}}= (\bn\cdot\delta \bx_{3.5\textrm{PN}})$ and 
\begin{align}
    \delta \bx_{3.5\textrm{PN}}^{i} = \frac{\delta m}{m}\biggl\{& \bigg[\frac{212G^3 m^3 \nu^2}{105 r^3} \dot{r}  + \frac{G^2 m^2 \nu ^2 }{r^2} \Big( \frac{78}{7} \dot{r} \bv^2 - \frac{26}{5} \dot{r}^3\Big)\bigg] \bx^{i} \nonumber\\
    + &\bigg[- \frac{8}{35} G m \bv^4 \nu ^2 - \frac{48 G^3 m^3 \nu ^2}{35 r^2} - \frac{G^2 m^2 \nu ^2}{r} \Big(\frac{58}{105} \bv^2 + \frac{398}{105} \dot{r}^2 \Big) \bigg]\bv^{i}\biggr\}
\end{align}
is the 3.5PN COM correction.

\subsection{Full result}
We now arrive at a full 4.5PN result by summing over all contributions as described above. The full result in relative coordinates can be written as
\begin{equation}
\ba^i_{4.5\textrm{PN}} = \mathcal{A} \bx^i + \mathcal{B} \bv^i,\label{eq:a4.5}
\end{equation}
where the coefficients $\mathcal A$ and $\mathcal B$ can be written as
\begin{align}
	\mathcal{A} &= \frac{G^5 m^5 \nu\dot{r}}{r^7}\biggl(- \frac{73178}{135} - \frac{56416}{105} \nu  - \frac{3040}{21} \nu ^2 \biggr) \\
	&+ \frac{G^4 m^4 \nu \dot{r} }{r^6} \biggl[ \biggl(\frac{493214}{315} + \frac{207052}{105} \nu  + \frac{9932}{35} \nu ^2 \biggr)\bv^2 - \biggl( \frac{2177438}{945} + \frac{1028644}{189} \nu  + \frac{96436}{105} \nu ^2 \biggr)\dot{r}^2 \biggr] \nonumber\\
	&+ \frac{G^3 m^3 \nu \dot{r} }{r^5} \biggl[ \biggl(- \frac{20747}{105} -1774 \nu  + \frac{3928}{3} \nu ^2 \biggr)\bv^4 + \biggl(- \frac{19151}{105} + \frac{108773}{15} \nu  - \frac{19756}{7} \nu ^2 \biggr)\bv^2\dot{r}^2 \nonumber\\
	&\qquad\quad + \biggl(\frac{21508}{21} - \frac{716027}{105} \nu  + \frac{83414}{105} \nu ^2 \biggr)\dot{r}^4 \biggr] \nonumber\\
	&+ \frac{G^2 m^2 \nu \dot{r} }{r^4} \biggl[ \biggl(- \frac{46372}{105} + \frac{74759}{35} \nu  - \frac{51532}{35} \nu ^2 \biggr)\bv^6 + \biggl(\frac{54085}{21} - \frac{258268}{21} \nu  + \frac{186635}{21} \nu ^2 \biggr)\bv^4 \dot{r}^2 \nonumber\\
	&\qquad\quad + \biggl(- \frac{12274}{3} + \frac{56683}{3} \nu  - \frac{40526}{3} \nu ^2 \biggr)\bv^2\dot{r}^4 + \big(1980 -8658 \nu  + 5823 \nu ^2 \big)\dot{r}^6 \biggr],\nonumber
\end{align}
and
\begin{align}
	\mathcal{B} &= \frac{G^5 m^5 \nu}{r^6}\biggl( \frac{927826}{2835} + \frac{73772}{105} \nu  + \frac{3008}{35} \nu ^2 \biggr) \\
	&+ \frac{G^4 m^4 \nu}{r^5}\biggl[ \biggl(- \frac{89926}{315} - \frac{305576}{315} \nu  - \frac{17972}{105} \nu ^2 \biggr)\bv^2 + \biggl(\frac{33926}{45} + \frac{13504}{3} \nu  + \frac{4380}{7} \nu ^2 \biggr)\dot{r}^2 \biggr] \nonumber\\
	&+ \frac{G^3 m^3 \nu}{r^4}\biggl[ \biggl(- \frac{673}{21} + \frac{173872}{315} \nu  - \frac{2376}{35} \nu ^2 \biggr)\bv^4 + \biggl(\frac{75211}{105} - \frac{1332301}{315} \nu  - \frac{79052}{105} \nu ^2 \biggr)\bv^2 \dot{r}^2 \nonumber\\
	&\qquad\quad + \biggl(- \frac{46224}{35} + \frac{177399}{35} \nu  + \frac{140618}{105} \nu ^2 \biggr)\dot{r}^4 \biggr] \nonumber\\
	&+ \frac{G^2 m^2 \nu}{r^3}\biggl[ \biggl(\frac{18124}{315} - \frac{9621}{35} \nu  + \frac{5004}{35} \nu ^2 \biggr)\bv^6 + \biggl(- \frac{23053}{35} + \frac{98632}{35} \nu  - \frac{39359}{35} \nu ^2 \biggr)\bv^4 \dot{r}^2  \nonumber\\
	& \qquad\quad + \biggl(\frac{28274}{21} - \frac{104015}{21} \nu  + \frac{19774}{21} \nu ^2 \biggr)\bv^2 \dot{r}^4 + \biggl(- \frac{2324}{3} + \frac{7070}{3} \nu  + \frac{833}{3} \nu ^2 \biggr)\dot{r}^6 \biggr]. \nonumber
\end{align}
The full result can also be found in the supplementary file.

\section{Consistency checks on results}\label{sec:checks}

\subsection{Quasicircular limit}
In this section, we consider the special case of quasicircular orbits. In particular, we will use the acceleration through 4.5PN order to compute the expression $\dot\omega/\omega^2$, where $\omega$ is a well-defined orbital frequency for the special case of quasicircular orbits. This expression, when written as a function of $\omega$, is gauge independent under a large class of gauge transformations and will allow comparison with the general results derived in \refdot\cite{bala1}. Additionally, one can use flux-balance arguments to compute $\dot\omega/\omega^2$ exclusively from the far-field, allowing a direct comparison between the near-field and far-field EFT regimes.

We follow the approach of \refdot\cite{bala1}. The quasicircular orbit limit is defined by the relations
\begin{align}
r \omega^2&=-\langle\mathbf{n} \cdot \ba\rangle, \\
v&=r \omega, \\
\dot{r}&=0 + \mathcal O(v^5).
\end{align}

Using these relations, and defining $\gamma\equiv Gm/r$, we can write
\begin{equation}
\frac{v^2}{r^2}=\omega^2=\gamma\biggl[1+\gamma(-3+\nu)+\gamma^{\!2}\biggl(\frac{41}{4} \nu+\nu^2\biggr)+\mathcal O(v^5)\biggr]
\label{eq:w2-of-gam}
\end{equation}
using the conservative equations of motion through 2PN, which can be inverted and written in terms of either $\omega$ or $v$ as
\begin{align}
    \gamma&=(Gm\omega)^{2/3}\biggl[1+\biggl(1-\frac{\nu}{3}\biggr)(Gm\omega)^{2/3}+\biggl(3-\frac{65 \nu}{12}\biggr)(Gm\omega)^{4/3}+\mathcal O(v^5)\biggr],\label{eq:gam_of_x}\\
    &=v^2\biggl[1+(3-\nu)v^2+\biggl(18-\frac{89}{4} \nu+\nu^2\biggr)v^4+\mathcal O(v^5)\label{eq:gam-of-v}
    \biggr].
\end{align}
Taking the circular limit of the radiation-reaction acceleration in Eq.~\eqref{eq:a4.5} and using Eq.~\eqref{eq:w2-of-gam}, we have 
\begin{equation}
    (\ba^i)^\mathrm{RR}_\mathrm{circ}=\frac{32 \gamma^4 \nu \bv^i}{5m^3}\biggl[1+\biggl(-\frac{3431}{336}+\frac54\nu\biggr)\gamma+\biggl(\frac{659217}{18144}+\frac{26095}{2016}\nu-\frac{7}{4}\nu^2\biggr)\gamma^2+\mathcal O(v^{10})\biggr].\label{eq:aRRcirc}
\end{equation}
Taking a time derivative of Eq.~\eqref{eq:gam-of-v} and solving for $\dot r$, reducing the acceleration terms using the conservative equations of motion through 2PN and nonconservative equations of motion in Eq.~\eqref{eq:aRRcirc}, we find
\begin{equation}
\dot{r}=-\frac{64}{5} \nu\gamma^3\biggl[1-\biggl(\frac{1751}{336}+\frac{7 \nu}{4}\biggr) \gamma
+\biggl(\frac{230879}{18144}+\frac{40981 \eta}{2016}+\frac{\eta^2}{2}\biggr)\gamma^2+\mathcal O(v^{10})\biggr].
\end{equation}

Finally, taking a time derivative of Eq.~\eqref{eq:w2-of-gam} and solving for $\dot\omega/\omega^2$ as a function of $\omega$, we find 
\begin{align}
\frac{\dot{\omega}}{\omega^2}&=\frac{96}{5} \nu(Gm \omega)^{5/3}\biggl[1-\biggl(\frac{743}{336}+\frac{11}{4} \nu\biggr)(Gm \omega)^{2/3}\notag
\\
&\qquad\qquad\qquad\qquad+\biggl(\frac{34103}{18144}+\frac{13661}{2016} \nu+\frac{59}{18} \nu^2\biggr)(Gm \omega)^{4/3}\biggr].
\end{align}
This exactly reproduces the near-field results of \refdot\cite{bala1} and far-field expression in \refdot\cite{Blanchet:1995ez}, a nontrivial check on our results.

\subsection{Energy flux-balance equation through NNLO}
In this section, we use the energy flux-balance equations as a consistency check on the radiative equations of motion. We expect the locally-induced power loss to be equivalent to the energy flux at infinity up to a total derivative that time-averages to zero. This total time derivative amounts to a redefinition of the local conserved energy, akin to a ``Schott'' term in electrodynamics, which vanishes in the far-field regime. Through NNLO, the energy flux-balance equation is given by
\begin{equation}
    \frac{dE}{dt}=- \biggl(\frac{G}{5}I_{i j}^{(3)} I_{i j}^{(3)} + \frac{G}{189} I_{i j k}^{(4)} I_{i j k}^{(4)} + \frac{16G}{45} J_{i j}^{(3)} J_{i j}^{(3)} + \frac{G}{84} J_{i j k}^{(4)} J_{i j k}^{(4) }+ \frac{G}{9072} I_{i j k l}^{(5)} I_{i j k l}^{(5)} +\ldots \biggr).
    \label{eq:dEdt}
\end{equation} 

We first show that the LO radiation-reaction acceleration is consistent with the energy flux-balance equation. With the LO radiation reaction ${\ba}^{i}_K = -2G/5 {\bx}_{K}^{j} I^{ij (5)}_{0}$, the 2.5PN energy flux-balance equation is given by
\begin{align}
	\frac{\ud E}{\ud t} &= -\frac{2G}{5} \sum_{A} m_A {\bx}^{i}_{A}{\bv}^{j}_{A} I^{ij(5)}_{0} = -\frac{G}{5} I^{ij(1)}_{0} I^{ij(5)}_{0},
\end{align}
which agrees with the mass quadrupole term in Eq.~\eqref{eq:dEdt} modulo a total time-derivative.

At NLO, the acceleration, as derived from the nonconservative Lagrangian, is given by
\begin{align}
		{\ba}^{i}_{A} = &-\frac{16G}{45}\epsilon^{ikl} \left({\bx}_A^{j}{\bx}_A^{k} J_{0}^{jl(6)} +3 {\bx}_{A}^{j} {\bv}_{A}^{k} J_{0}^{j l(5)} \right) + \frac{G}{63}{\bx}_A^{j}{\bx}_A^{k}I^{ijk(7)}_{0}\nonumber\\
		& -\frac{2G}{5}{\bx}_{A}^{j}I^{ij(5)}_{1} -\frac{G}{5}\biggl(\frac{\partial I^{ij}_{1-}}{\partial {\bx}_{A-}^{i}}I^{ij(5)}_{0+} - \frac{\ud }{\ud t} \frac{\partial I^{ij}_{1-}}{\partial {\bv}_{A-}^{i}}I^{ij (5)}_{0+}\biggr)_{\mathrm{PL}}  .
\end{align}
We do not need to consider the order-reduced accelerations arising from substituting the 2.5PN acceleration in the 1PN conservative acceleration; these do not contribute to energy loss due to energy conservation. Integrating by parts liberally, for example with $I^{(6)}_{ij,0}$ and $I^{(7)}_{ij,0}$, the NLO energy flux becomes
\begin{align}
	\frac{\ud E}{\ud t} &= \frac{16G}{45}J^{}_{jl,0} J_{ij,0}^{(6)} + \frac{G}{189}I^{(1)}_{ijk,0}I^{(7)}_{ijk,0} -\frac{G}{5} I^{(1)}_{ij,0} I^{(5)}_{ij,1} -\frac{G}{5} I^{(1)}_{ij,1} I^{(5)}_{ij,0} ,
\end{align}
which again agrees with the energy-flux balance equation, Eq.~\eqref{eq:dEdt}, at NLO modulo a total time-derivative.

Similarly, given the NNLO accelerations in sections \ref{sec:Amq}--\ref{sec:Aco} (again neglecting the reduced conservative equations of motion) we find that 
\begin{align}
\frac{\ud E}{\ud t} & =-\frac{G}{84}\sum_{A}m_{A}\epsilon^{ijk}{\bv}_{A}^{i}{\bx}_{A}^{k}{\bx}_{A}^{l}{\bx}_{A}^{m}J_{0}^{jlm(8)}-\frac{G}{2268}\sum_{A}m_{A}{\bv}_{A}^{i}{\bx}_{A}^{j}{\bx}_{A}^{k}{\bx}_{A}^{l}I_{0}^{ijkl(9)}\nonumber\\
 & +\frac{16G}{45}J_{0}^{jl}J_{1}^{ij(6)}+\frac{G}{189}I_{0}^{ijk(1)}I_{1}^{ijk(7)}-\frac{G}{5}I_{0}^{ij(1)}I_{i2}^{j(5)}-\frac{G}{5}I_{1}^{ij(1)}I_{1}^{ij(5)}\nonumber\\
 & +\sum_{A}{\bv}_{A}^{i}\biggl[-\frac{G}{5}\biggl(\frac{\partial I_{2-}^{jk}}{\partial{\bx}_{A-}^{i}}I_{0+}^{jk(5)}-\frac{\ud}{\ud t}\frac{\partial I_{2-}^{ij}}{\partial{\bv}_{A-}^{i}}I_{0+}^{ij(5)}\biggr)-\frac{16G}{45}\biggl(\frac{\partial J_{1-}^{ij}}{\partial{\bx}_{A-}^{i}}J_{0+}^{ij(5)}-\frac{\ud}{\ud t}\frac{\partial J_{1-}^{ij}}{\partial{\bv}_{A-}^{i}}J_{0+}^{ij(5)}\biggr)\nonumber\\
 &\qquad+\frac{G}{189}\biggl(\frac{\partial I_{1,-}^{ijk}}{\partial{\bx}_{A-}^{i}}I_{0+}^{ijk(7)}-\frac{\ud}{\ud t}\frac{\partial I_{1-}^{ijk}}{\partial{\bv}_{A-}^{i}}I_{0+}^{ijk(7)}\biggr)\biggr]_{\mathrm{PL}}.
\end{align}
To show that this is consistent with the right-hand side of Eq.~\eqref{eq:dEdt}, we must simplify this expression. We collect the higher time-derivative terms, explicitly calculate the variations and rewrite them in terms of multipole moments. This is a lengthy but straightforward process;
it can be shown that left-hand side of the NNLO energy flux then becomes 
\begin{align}
	\frac{\ud E}{\ud t} &= -\frac{G}{84} J_{ijk,0} J_{ijk,0}^{(8)}   -\frac{G}{9072}I_{ijkl,0}^{(1)} I_{ijkl,0}^{(9)} +\frac{16G}{45}\left(J_{jl,0} J_{ij,1}^{(6)}+J_{jl,1} J_{ij,0}^{(6)}  \right) \nonumber\\
	&+ \frac{G}{189}\left(I^{(1)}_{ijk,0}I^{(7)}_{ijk,1} +I^{(1)}_{ijk,1}I^{(7)}_{ijk,0}\right) -\frac{G}{5} \left(I^{(1)}_{ij,0} I^{(5)}_{ij,2} + I^{(1)}_{ij,1} I^{(5)}_{ij,1} + I^{(1)}_{ij,2} I^{(5)}_{ij,0}\right),
\end{align}
which again agrees the energy flux-balance equation, Eq.~\eqref{eq:dEdt}, at NNLO modulo a total time derivative. This check helps establish the validity of the multipole moments in use and the nonconservative action approach for the derivation of the radiation-reaction effects.

\section{Conclusion}\label{sec:conclude}
In this paper, we  calculate the radiation reaction force at the 4.5PN order for non-spinning compact binary inspiral completely in the EFT approach. This amounts to a 2PN correction to the leading Burke--Thorne radiation-reaction term. To accomplish this, we used the recent EFT calculation of the 2PN correction to the mass quadrupole, calculated in \refdot\cite{radnrgr}.

We can write the results in terms of the acceleration of one of the binary constituents or in terms of the relative coordinates. In order to write in terms of the relative coordinates, we need to transform into the center-of-mass frame. For the 4.5PN result, we need to include the 3.5PN radiative correction to the center-of-mass, which is nonzero in our chosen gauge. We calculate this correction to the center-of-mass in the appendix, and then use this to write the results in terms of the relative coordinates.

As a first consistency check, we calculated the adiabatic parameter $\dot\omega/\omega^2$ 
in the quasicircular limit. Since this expression is gauge independent under a large class of gauge transformations, we can compare to other calculations and find agreement with the results presented in \refdot\cite{bala1}. As a second consistency check, we calculate the flux-balance equation at NNLO and find agreement up to a total time derivative, as expected. 

Combining the results in this paper with previous results, we have  completed the nonconservative corrections to the equations of motion for spinning compact binaries through 4.5PN order derived entirely using the EFT approach. This is an important step towards completing the equations of motion through 5PN for use in producing templates for gravitational wave detectors now and in the future.

\section{Acknowledgements}

A.K.L.\ and B.P. are supported in part by the National Science Foundation under Grant No.~PHY-2112829. Z.Y.\ is supported by ERC Consolidator Grant ``Precision Gravity: From the LHC to LISA,'' provided by the European Research Council (ERC) under the European Union's H2020 research and innovation programme, grant No.817791. We are grateful for all the early work on this project from Ira Rothstein and Nat\'alia Maia. We acknowledge extensive use of the xAct packages \cite{xact}.
%======================

\appendix
\section{Radiative center-of-mass corrections} \label{sec:appendix}
Energy and momentum loss due to gravitational radiation emission leads to a shift in the COM momentum and position of a binary system. In our gauge, the leading radiative corrections to the COM are 3.5PN effects, which enter the PN equations of motion at 4.5PN. 

The conservative definition of the COM vector $\bG^i$ relates to the linear momentum~$\bP^i$ through the Noetherian integral $\bK^i = \bG^i -t \bP^i $. The invariance of the conservative Lagrangian under the Lorentz boost leads to the conservation of $\bK^i$, which implies that \cite{Blanchet:2002mb}
	\begin{equation}
	\label{eq:Noether}
		\frac{\ud \bG^i}{\ud t} = \bP^i,
	\end{equation}
where $\bP^i$ remains constant, i.e.,
\begin{equation}
\label{eq:Pconserv}
	\frac{\ud \bP^i}{\ud t} = 0
\end{equation}	
for all conservative PN orders. This relation no longer holds upon the inclusion of dissipative effects. With radiation reaction considered in the binary dynamics, the flux-balance equations for the energy, angular momentum, and linear momentum can be combined to compute the averaged secular evolution of the binaries. In this section, we focus on solving for a COM position $\bG^i$ at 3.5PN that is consistent with the flux-balance equation results in \refdot\cite{Blanchet:1996vx}. 

At the leading 2.5PN order, the balance equation for total linear momentum including the net force can be written as
\begin{align}
\label{eq:dP2.5}
	\sum_A m_A {\ba}_{A, 2.5\textrm{PN}}^{i}
 + \frac{\ud \bP^i_{2.5\textrm{PN}}}{\ud t}\bigg|_{a_{0\textrm{PN}}} 
 = -\frac{2G}{5}I^{j}I^{ij(5)},
\end{align}
where the mass-type dipole moment $I^i = \int d^3 \bx T^{00} \bx^i$ corresponds to the conserved COM $\bG^i$. The $\bP^i_{2.5\textrm{PN}}$ is a possible linear momentum term at 2.5PN that can be solved by the equation above, similar to the ``Schott'' term in electromagnetism \cite{schott}. The Schott-like momentum depends on the expression of the radiation-reaction force on the right-hand side of Eq.~\eqref{eq:dP2.5}. We can rewrite some of the time derivatives on the multipole moments with the addition of a total time derivative that can be absorbed into the left-hand side, equivalent to a gauge transformation. 

Using the Burke--Thorne acceleration \eqref{eq:aBT} and the leading mass dipole $I^i = \sum_A m_A {\bx}_A^i$, it can be shown from  Eq.~(\ref{eq:dP2.5}) that there is no net radiation effects on the linear momentum at 2.5PN 
related by some gauge transformation, i.e., $\bP^i_{2.5\textrm{PN}}=0$.
With the 3.5PN radiation reaction included, the balance equation for linear momentum is given by \cite{Blanchet:1996vx} 
\begin{align}
\label{eq:Pi_Blanchet}
	\frac{\ud \bP^i}{\ud t} = - \frac{2G}{5}I_{j}I_{ij}^{(5)} -\left(\frac{2G}{63 } I_{i j k}^{(4)} I_{j k}^{(3)}+\frac{16G}{45} \epsilon_{i j k} I_{j m}^{(3)} J_{k m}^{(3)} \right) + \mathcal{O}\left(\epsilon^{4.5}\right),
\end{align}
where the net force at 3.5PN contains contributions from
\begin{align}
\label{eq:dP3.5}
	\left.\frac{\ud \bP^i}{\ud t}\right|_{3.5\textrm{PN}} = \sum_A m_A {\ba}_{A, 3.5\textrm{PN}}^{i} + \frac{\ud \bP^i_{1\textrm{PN}}}{\ud t}\bigg|_{\ba_{2.5\textrm{PN}}} + \frac{\ud \bP^i_{3.5\textrm{PN}}}{\ud t} \bigg|_{\ba_{0\textrm{PN}}},
\end{align}
with $\bP^i_{3.5\textrm{PN}}$ a possible Schott term modification to the linear momentum at 3.5PN. Equating Eq.~\eqref{eq:Pi_Blanchet} and Eq.~\eqref{eq:dP3.5} to solve for $\bP^i_{3.5\textrm{PN}}$ leads to an explicit expression for the secularly evolving linear momentum consistent with the flux-balance equations. At 3.5PN order, Eq.~\eqref{eq:Pi_Blanchet} includes
\begin{align}
\label{eq:Pi_Blanchet_3.5}
	\frac{\ud \bP^i}{\ud t}\bigg|_{3.5\textrm{PN}}\!\!\!\! =& - \frac{2G}{5}I_{j,0}I^{(5)}_{ij,1}- \frac{2G}{5}I_{j,1}I^{(5)}_{ij,0} - \frac{2G}{5}I_{j,0}I^{(5)}_{ij,0} \bigg|_{\ba_{1\textrm{PN}}} \!\!\!\!- \frac{2G}{63 } I_{i j k}^{(4)} I_{j k}^{(3)} - \frac{16G}{45} \epsilon_{i j k} I_{j m}^{(3)} J_{k m}^{(3)},
\end{align}
with the LO mass octupoles and current quadrupoles. For the terms in Eq.~\eqref{eq:dP3.5}, the radiation-reaction force $\sum_A m_A {\ba}_{A, 3.5\textrm{PN}}^{i}$ taken from \refdot\cite{PhysRevD.86.044029} includes
\begin{align}
\label{eq:P0PN_a3.5PN}
	\sum_A m_A {\ba}_{A, 3.5\textrm{PN}}^{i}=& \sum_A \frac{\delta}{\delta {\bx}_{A-}^{i}}\biggl(-\frac{G}{5} I_{0-}^{ij} I_{1+}^{ij(5)} -\frac{G}{5} I_{1-}^{ij} I_{0+}^{ij(5)} -\frac{16G}{45} J_{0-}^{ij} J_{0+}^{ij(5)} + \frac{G}{189} I_{0-}^{ijk} I_{0+}^{ijk(7)} \biggr)_{\textrm{PL}} \nonumber \\
	&-\frac{2G}{5} m_A {\bx}^{j}_A I_{ij}^{(5)} \bigg|_{\ba_{1\textrm{PN}}} + \biggl(\frac{\partial \cL_{1\textrm{PN}}}{\partial {\bx}^i_{A}} - \frac{\ud}{\ud t} \frac{\partial \cL_{1\textrm{PN}}}{\partial {\bv}^i_{A}}   \biggr)\bigg|_{\ba_{2.5\textrm{PN}}},
\end{align}
where $L_\text{1PN}$ is the 1PN conservative Lagrangian, and the 1PN momentum can be derived from
\begin{align}
\label{eq:P1PN_a2.5PN}
	\frac{\ud \bP^i_{1\textrm{PN}}}{\ud t}\bigg|_{\ba_{2.5\textrm{PN}}} = \frac{\ud}{\ud t}\biggl(\sum_A \frac{\partial \cL_{1\textrm{PN}}}{\partial {\bv}^i_{A}}  \biggr)\bigg|_{\ba_{2.5\textrm{PN}}},
\end{align}
with $\sum_A \partial \cL_{1\textrm{PN}}/\partial {\bx}^i_{A} =0$. Cancellations from Eq.~\eqref{eq:dP3.5} and Eq.~\eqref{eq:Pi_Blanchet_3.5} give
\begin{align}
\label{eq:cancel_1}
	&\sum_A \frac{\delta}{\delta {\bx}_{A,-}^{i}}\left( -\frac{G}{5} I_{1-}^{ij} I_{0+}^{ij(5)} -\frac{16G}{45} J_{0-}^{ij} J_{0+}^{ij(5)} + \frac{G}{189} I_{0-}^{ijk} I_{0+}^{ijk(7)} \right)_{\textrm{PL}}+ \frac{\ud \bP^i_{3.5\textrm{PN}}}{\ud t}  \nonumber \\
	=& - \frac{2G}{5}I^j_{1}I^{ij(5)}_{0} - \frac{2G}{63 } I_{0}^{i j k(4)} I_{0}^{j k(3)} - \frac{16G}{45} \epsilon^{i j k} I_{0}^{j m(3)} J_{0}^{k m(3)}.
\end{align}

Integrating by parts liberally and performing the variation of the minus coordinates, we find
{
\allowdisplaybreaks
\begin{align}
	\frac{\ud \bP^i_{3.5\textrm{PN}}}{\ud t} = &\frac{\ud }{\ud t}\biggl\{\frac{G}{63}(-I_{ijk}^{(6)}I_{jk} + I_{ijk}^{(5)}I_{jk}^{(1)} - I_{ijk}^{(4)}I_{jk}^{(2)} - I_{ijk}^{(3)}I_{jk}^{(3)} + I_{ijk}^{(2)}I_{jk}^{(4)} )
 \nonumber\\
 &\quad+ \frac{8G}{45}\epsilon_{ijk}(2I_{jl}J_{kl}^{(5)} + I_{jl}^{(1)}J_{kl}^{(4)} -  I_{jl}^{(2)}J_{kl}^{(3)} -  I_{jl}^{(3)}J_{kl}^{(2)} +  I_{jl}^{(4)}J_{kl}^{(1)})+ \frac{8G}{15} J_{j}J_{ij}^{(4)}	  
 \nonumber\\
&\quad + \frac{G}{105}\sum_A m_A\Bigl[(11 \bx_{A}^{2}{\bx}_{A}^{j}\delta^{ik} -17{\bx}_{A}^{i}{\bx}_{A}^{j}{\bx}_{A}^{k} )I_{jk}^{(6)} 
\nonumber\\
&\qquad
 + \bigl(34 (\bx_{A}\cdot\bv_{A})
 {\bx}_{A}^{j}\delta^{ik}-11 \bx_{A}^{2}{\bv}_{A}^{j}\delta^{ik} -46{\bv}_{A}^{i}{\bx}_{A}^{j}{\bx}_{A}^{k} - 22{\bx}_{A}^{i}{\bx}_{A}^{j}{\bv}_{A}^{k} \bigr)I_{jk}^{(5)}\Bigr]\biggr\} 
 \nonumber\\
	&-\sum_A \frac{8G}{15}m_A \epsilon_{jkl}{\bx}_{A}^{j}{\ba}_{A}^{k} J_{il}^{(4)	}-I_{jk}^{(5)}\biggl[ \frac{G}{63}I_{ijk}^{(2)} + \frac{8G}{45} \epsilon_{ijl}J_{kl}^{(1)} \nonumber\\
	&+ \frac{G}{105}\sum_{A} m_A \biggl(-22 {\bx}_{A}^{i}{\bv}_{A}^{j}{\bv}_{A}^{k} - 22{\bx}_{A}^{i}{\bx}_{A}^{j}{\ba}_{A}^{k} + 12 {\bv}_{A}^{i}{\bx}_{A}^{j}{\bv}_{A}^{k} + 17 {\ba}_{A}^{i}{\bx}_{A}^{j}{\bx}_{A}^{k} 	\nonumber\\
	&\quad+\delta^{ik}\Bigl(8\bv_{A}^2 {\bx}_A^{j} + 34 (\bx_{A}\cdot \ba_{A}){\bx}_{A}^{j} + 12(\bx_{A}\cdot \bv_{A}){\bv}_{A}^{j} - 11\bx_{A}^{2}{\ba}_{A}^{j} + \frac{21Gm_B {\bx}_{A}^{j}}{r} \Bigr)\biggr) \biggr],
\end{align}
}in which the terms outside the total time derivative vanish after substituting the Newtonian equations of motion and the LO multipole moments. Therefore an explicit linear momentum $P^i$ that obeys the flux-balance equation at the 3.5PN order is given by
\begin{align}
	\bP^i_{3.5\textrm{PN}} & = \frac{G}{63}(-I_{ijk}^{(6)}I_{jk} + I_{ijk}^{(5)}I_{jk}^{(1)} - I_{ijk}^{(4)}I_{jk}^{(2)} - I_{ijk}^{(3)}I_{jk}^{(3)} + I_{ijk}^{(2)}I_{jk}^{(4)} )
 \nonumber\\
 &\quad+ \frac{8G}{45}\epsilon_{ijk}(2I_{jl}J_{kl}^{(5)} + I_{jl}^{(1)}J_{kl}^{(4)} -  I_{jl}^{(2)}J_{kl}^{(3)} -  I_{jl}^{(3)}J_{kl}^{(2)} +  I_{jl}^{(4)}J_{kl}^{(1)})+ \frac{8G}{15} J_{j}J_{ij}^{(4)}	  
 \nonumber\\
&\quad + \frac{G}{105}\sum_A m_A\Bigl[(11 \bx_{A}^{2}{\bx}_{A}^{j}\delta^{ik} -17{\bx}_{A}^{i}{\bx}_{A}^{j}{\bx}_{A}^{k} )I_{jk}^{(6)} 
\nonumber\\
&\qquad
 + \bigl(34 (\bx_{A}\cdot\bv_{A})
 {\bx}_{A}^{j}\delta^{ik}-11 \bx_{A}^{2}{\bv}_{A}^{j}\delta^{ik} -46{\bv}_{A}^{i}{\bx}_{A}^{j}{\bx}_{A}^{k} - 22{\bx}_{A}^{i}{\bx}_{A}^{j}{\bv}_{A}^{k} \bigr)I_{jk}^{(5)}\Bigr].
\end{align}

Next, the 3.5PN COM position $\bG^{i}$ is related to $\bP^i_{3.5\textrm{PN}}$ by \cite{Blanchet:2018yqa}
\begin{align}
\label{eq:Gi_Blanchet_3.5}
	\frac{\ud \bG^{i}}{\ud t} = \bP^{i} - \frac{2G}{21} I_{}^{ijk(3)}I_{}^{jk(3)}.
\end{align}
We equate Eq.~\eqref{eq:Gi_Blanchet_3.5} with the total 3.5PN expansion of the flux of the COM position, which can be constructed by some total time derivatives, 
\begin{align}
\frac{\ud \bG^{i}}{\ud t} &= \frac{\ud \bG^{i}_{3.5\textrm{PN}}}{\ud t} + \frac{\ud \bG^{i}_{1\textrm{PN}}}{\ud t} \bigg|_{\ba_{2.5\textrm{PN}}} \nonumber\\
	&= \frac{\ud}{\ud t} \biggl\{ \frac{G}{63}(-I_{jk}I_{ijk}^{(5)}+ 2I_{jk}^{(1)}I_{ijk}^{(4)}-3I_{jk}^{(2)}I_{ijk}^{(3)} -4I_{jk}^{(3)}I_{ijk}^{(2)}+5I_{jk}^{(4)}I_{ijk}^{(1)}) \nonumber\\
	&\quad+ \frac{8G}{45}\epsilon^{ijk}(2I_{jm}J_{km}^{(4)} - I_{jm}^{(1)}J_{km}^{(3)}  - I_{jm}^{(3)}J_{km}^{(1)} + 2I_{jm}^{(4)}J_{km}) + \frac{8G}{15}J^{k}J_{}^{ik(3)} \nonumber\\
	&\quad+\frac{G}{105}I_{kl}^{(5)} \sum_{A } m_A (11 \delta^{ik}\bx^{2}_{A} \bx_{A}^{l} - 17\bx_{A}^{i}\bx_{A}^{l}\bx_{A}^{k} ) \biggr\} - \frac{8G}{15}\epsilon^{jkl}\sum_A m_{A}\bx_{A}^{k} \ba_{A}^{l} J_{}^{ij(3)},
	\label{eq:G3.5final}
	\end{align}
 where all multipole moments are their LO expressions.
The last term outside the total derivative vanishes after substituting the Newtonian equations of motion. Therefore, the COM position $\bG^i$ at 3.5PN can be determined as the Schott terms inside the time derivative of Eq.~\eqref{eq:G3.5final}, which contributes to a 4.5PN piece of corrections to the 1PN acceleration.

%===================================

\bibliographystyle{unsrt}
\bibliography{Ref45BT}

\end{document}